\documentclass[aps,twocolumn,showpacs,amsmath,amssymb]{revtex4}
\usepackage{dcolumn}
\usepackage{bm}
\usepackage{color}
\usepackage{latexsym}
\usepackage{amssymb}
\usepackage{graphicx}
\usepackage{ulem}

\begin{document}
\title{Quantum capacitance of the monolayer graphene}

\author{M. V. Cheremisin}

\affiliation{A.F.Ioffe Physical-Technical Institute, 194021
St.Petersburg, Russia}

\begin{abstract}
The quantum capacitance of the monolayer graphene for arbitrary carrier density, magnetic field, temperature and LL broadening is found. The density dependence of the quantum capacitance is analyzed when magnetic field(temperature) is fixed(varied) and vice versa. High-field induced splitting of the LL energy spectrum is examined. The theory is compared with experimental data.
\end{abstract}


\maketitle

Recently, a great deal of interest has been focused on the electric field effect and transport in a two-dimensional electron gas system formed in graphene flake \cite{Novoselov04}. In the present paper, we are are mostly concerned with the quantum capacitance of the monolayer graphene placed in magnetic field.

\begin{figure}
\begin{center}
\includegraphics[scale=0.75]{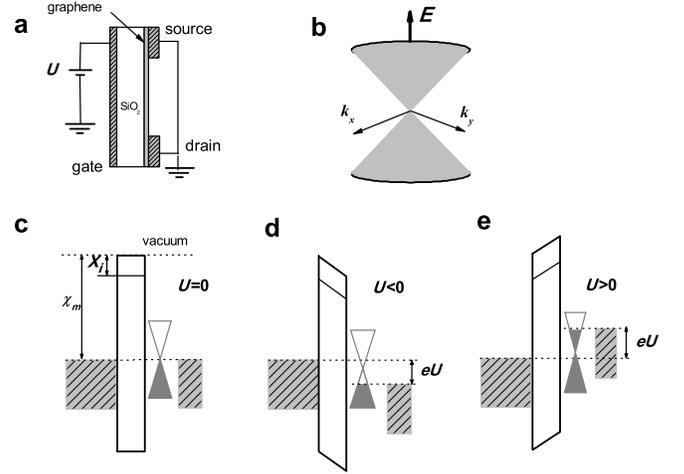} \caption[]{\label{Fig1}(a)Experimental setup for gated graphene.
(b)Band structure at $\mathbf{k} \simeq 0$. The energy diagram for (c)$U=0$,
(d)$U<0$ and (e)$U>0$, where $\chi_{m}$ is the metal work function, and $X_{i}$ the electron affinity of the insulator.}
\end{center}
\end{figure}

The typical experimental setup is shown in Fig.\ref{Fig1}, a. The metal backgate and the graphene flake are connected via the source of voltage, $U$, serving to change the carrier density. According to Refs.\cite{Wallace47},\cite{McClure56}, the energy spectrum obeys the linear dependence
\begin{equation}
E(k)=\hbar \upsilon k, \label{SLG spectrum_B0}
\end{equation}
where $E(k)$ is the energy, and $k$, the distance in the $\mathbf{k}$-space relative to the zone edge ( see Fig.\ref{Fig1}, b ), $\upsilon$ is Fermi velocity. The $k>0(k<0)$ refers to the electron(hole) conducting  bands, respectively. The state $E=0$ is called the Dirac point(DP). It will be remind that the Fermi energy, $\mu$, in graphene can be varied either by means of backgate voltage via field-effect\cite{Novoselov04} or chemical doping. When $\mu > 0$ ($\mu < 0$), the Fermi level falls in the electron ( hole ) conducting band, respectively. For $\mu=0$ the Fermi level coincides with the Dirac point, the density of the conducting electrons is equal to that of holes.

Following \cite{Kim12}, we plot in Figs.\ref{Fig1}, c-e the energy diagram for an arbitrary gate voltage bias. The applied gate voltage $U$ consists of the voltage drop across the capacitance and the voltage associated with the Fermi level of the graphene:
\begin{equation}
U= Q/C+\mu/e, \label{Field-Effect equation}
\end{equation}
where $Q$ is the charge density of the graphene monolayer, $C=\epsilon_{0}\epsilon /d$ is the capacitance per unit area, $d$ is the gate thickness, $\epsilon_{0}$ and, $\epsilon$, are the permittivity of free space and the relative permittivity of the substrate, respectively. It is to be noted that Eq.(\ref{Field-Effect equation}) gives the total capacitance $C_{tot}=\frac{dQ}{dU}$ of the graphene structure as $1/C_{tot}=1/C+1/C_{q}$, where $C_{q}=e\frac{dQ}{d\mu}$ is the so-called\cite{Luryi88,Smith85} the quantum capacitance. Usually, the condition $C \ll C_{q}$ is fulfilled, therefore the charge density in the graphene monolayer yields the simple relationship $Q=CU$ known within conventional field-effect formalism. Further, we discuss the validity of the field-effect approach.

In general, the graphene can exhibit the charge-neutrality state $Q=0$ at a certain Fermi energy. Without chemical doping, the neutrality occurs at the Dirac point. For simplicity, we further neglect the chemical doping.

Using the Gibbs statistics, we can distinguish the components of the thermodynamic potential for electrons,
$\Omega_{e}$, and holes, $\Omega_{h}$, and, then represent\cite{Cheremisin11} them as follows
\begin{eqnarray}
\Omega_{e} =-kT\sum \limits_{k} \ln \left(1+e^{\frac{\mu -|E(k)|
}{kT}}\right),
\label{SLG_OMEGA}\\
\Omega_{h}=\Omega_{e}(-\mu),
\nonumber
\end{eqnarray}
which gives the electron (hole) concentration as:
\begin{equation}
N=- {\partial \Omega_{e}
\overwithdelims()\partial \mu }_{T}, P={\partial \Omega_{h}
\overwithdelims()\partial \mu }_{T}.
\label{NP_DEFINITION}
\end{equation}

\begin{figure}
\begin{center}
\includegraphics[scale=0.75]{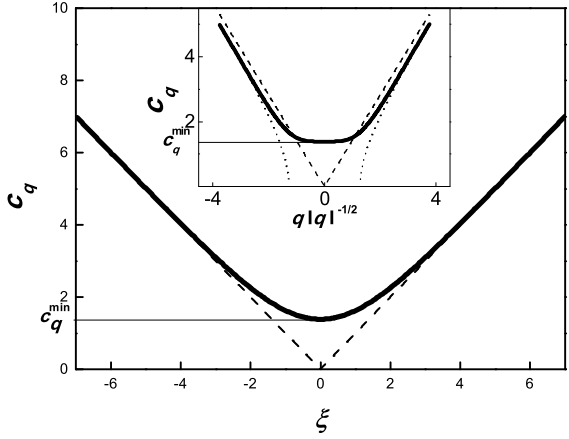} \caption[]{\label{Fig2} Dimensionless quantum capacitance $c_{q}$ vs reduced Fermi energy $\xi$(main panel) and variable $q|q|^{-1/2}$(inset), where $q=Q/eN_{T}$ is the dimensionless charge density. The dotted and dashed asymptotes correspond to intermediate $|\xi| \geq 1$ and strong $|\xi| \gg 1$ degeneracy case respectively. The horizontal line visualizes the quantum capacitance minimum $c_{q}^{min}=2\ln2$ at the Dirac point.}
\end{center}
\end{figure}

Using the graphene density of states(DOS), $D_{0}(E)=\frac{2|E|}{\pi \hbar ^{2}\upsilon^{2}}$, which includes both
the valley and the spin degeneracies, we obtain
\begin{equation}
N=N_{T}F_{1}(\xi ), P=N(-\xi ),
\label{SLG_N_B0}
\end{equation}
where $\xi =\mu/kT$ is the degeneracy parameter, $F _{n}(z)$ is the Fermi integral and, $N_{T}=\frac{2}{\pi}\left( \frac{kT}{\hbar \upsilon}\right )^{2}$. For two opposite cases of strong $\xi \geq 1$ and weak $\xi \ll 1$  degeneracy the electron density yields
\begin{eqnarray}
N=N_{T}\left (\frac{\xi^{2}}{2}+\frac{\pi^{2}}{6} \right ), \xi \geq 1
\label{N_strong_degeneracy} \\
N=N_{T}\left ( \frac{\pi^{2}}{12}+\xi \ln2+\frac{\xi^{2}}{4} \right ), \xi \ll 1.
\label{N_weak_degeneracy}
\end{eqnarray}
At $T=0$ Eq.(\ref{N_strong_degeneracy}) gives the density of degenerate electrons as $N_{0}=\frac{1}{\pi}\left( \frac{\mu}{\hbar \upsilon}\right )^{2}$.
With the help of Eq.(\ref{SLG_N_B0}) one can easily investigate the hole carriers case as well.

\section{ \label{sec: Quantum capacitance at zero magnetic field}Quantum capacitance at zero magnetic field}

Let us calculate the quantum capacitance based on the definition $C_{q}=e\frac{dQ}{d\mu}$, where $Q=e(N-P)$ is the total charge density. With the help of Eq.(\ref{SLG_N_B0}) we reproduce the result reported in Ref.\cite{Fang07} as
\begin{equation}
C_{q}=C_{0}\ln \left[2(1+\cosh\xi)\right],
\label{capacitance} \\
\end{equation}
where $C_{0}=e^{2}\frac{D(\mu)}{\xi}=\frac{e^{2}N_{T}}{kT}$ is the  dimensional unit of the capacitance. In Fig.\ref{Fig2},main panel we plot the dependence of the dimensionless quantum capacitance $c_{q}=C_{q}/C_{0}$ vs reduced Fermi energy $\xi$. The dependence is V-shaped. For degenerated carriers $|\xi| \gg 1$ the quantum capacitance obeys the linear asymptote $c_{q}=\xi$ shown by the dashed line in Fig.\ref{Fig2}. Then, in the vicinity of the Dirac point $|\xi| \ll 1$ the Eq.(\ref{capacitance}) provides the capacitance minimum as $c_{q}^{min}=2\ln{2}$.

\begin{figure}
\begin{center}
\includegraphics[scale=0.75]{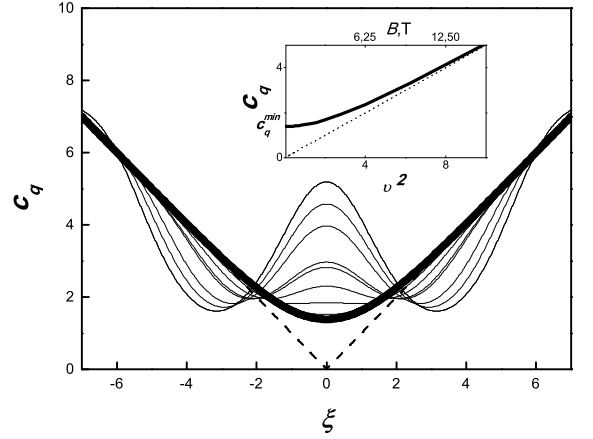} \caption[]{\label{Fig3} Dimensionless quantum capacitance vs reduced Fermi energy at fixed $T$=250K for $B$=0(bold line) and $B$=2,4..16 � ($\vartheta=1.1-3.2$). The dashed line depicts the degenerated carriers asymptote at T,$B$=0. Inset: quantum capacitance at the zero-charge point $\xi=0$ vs dimensionless magnetic field $\vartheta^{2}=\pi \Gamma \left(\frac{\hbar \upsilon}{kT} \right)^{2}$. Dotted line corresponds to high-B limit.}
\end{center}
\end{figure}

Overwise, we can represent the quantum capacitance $c_{q}$ as a function of the charge density itself\cite{Ponomarenko10} since the latter is proportional to the gate voltage. With the help of Eq.(\ref{SLG_N_B0}) in Fig.\ref{Fig2}, inset we plot the quantum capacitance vs variable $q|q|^{-1/2}$, where $q=Q/eN_{T}$ is the dimensionless charge density. In the high-degeneracy limit $|\xi| \gg 1$ the quantum capacitance follows the linear asymptote $c_{q}=\sqrt{2|q|}$ shown in Fig.\ref{Fig2},inset by the dashed line. However, for intermediate degeneracies $|\xi|\geq 1$ the quantum capacitance obeys somewhat different asymptote $c_{q}=\sqrt{2(|q|-\pi^{2}/6)}$ represented by the dotted line. The interchange between both asymptotes is caused by the degeneracy assisted change in the carrier density dependence specified by Eq.(\ref{N_strong_degeneracy}). Then, for non-degenerated carriers $|\xi| \ll 1$ the dimensionless quantum capacitance remains nearly constant, i.e. $c_{q} \sim c_{q}^{min}$ within wide range of the charge densities $|q| \leq q_{cr}=2\ln^{2}2$.

Let us now estimate the typical values of the quantum capacitance. For bath temperature T=10K and velocity $\upsilon=1.15 \times 10^{8}$ cm/s, reported in Ref.\cite{Ponomarenko10} we calculate the capacitance minimum $C_{q}^{min}=c_{q}^{min}C_{0}$=21 nF/cm$^{2}$ and the critical electron density $N_{cr}=q_{cr}N_{T}=7.7 \times 10^{7}$cm$^{-2}$. These values are two orders of magnitude lower compared to respective experimental values $C_{q}^{min}$=1 $\mu$F/cm$^{2}$ and $N_{cr}=4 \times 10^{11}$cm$^{-2}$. The aforementioned disagreement between theory expectation and experimental observations can be attributed to formation of the electron-hole plasma puddles argued \cite{Xu11},\cite{Mayorov12} to persist in graphene at the Dirac point. The experimental data\cite{Ponomarenko10} confirm the above suggestion since the quantum capacitance minimum remains unchanged up to $T=250$K.

We now intend to verify the validity of the field-effect relationship $U=Q/C$ for actual graphene systems. For usual thickness $\sim300$nm of SiO$_{2}$ gate dielectric the geometric capacitance yields $C=1.17$nF/cm$^{2}$. Field-effect approach remains valid when $C\ll C_{q}$. The above inequality could be violated initially at zero-charge point where the quantum capacitance exhibits the minimum value. Arguing that the quantum capacitance depends on temperature $C^{min}_{q} \sim T$, we estimate the validity of field-effect approach as $T>5$K. At lower temperatures the carrier charge density can be found rigorously on the basis of Eq.(\ref{Field-Effect equation}). In real systems, the quantum capacitance at the DP can be, however, strongly enhanced because of the electron-hole puddles\cite{Xu11}\cite{Mayorov12}, therefore the condition  $C\ll C_{q}$ is always fulfilled.

\section{ \label{sec: Quantum capacitance in magnetic field}Quantum capacitance in presence of the magnetic field}

We now attempt to find out the quantum capacitance of the monolayer graphene placed in magnetic field. The energy spectrum is given by\cite{McClure56}:
\begin{equation}
E_{\mathcal{N}}= \pm \frac{\hbar \upsilon}{\lambda_{B}}\sqrt{2\mathcal{N}},
\label{SLG_SPECTRUM_B}
\end{equation}
where $\lambda_{B}=\sqrt{\hbar c/eB}$ is the magnetic length, and $\mathcal{N}=0,1,2..$, the Landau level(LL)
number. Here, the $\pm$ sign refers to the electron(+) and hole(-) energy bands, respectively.
For a moment, we assume no LL broadening.

With the energy spectrum specified by Eq.(\ref{SLG_SPECTRUM_B}), the electron and the hole
part of the thermodynamic potential can be written as it follows
\begin{eqnarray}
\Omega_{e}=-kT \Gamma \left[4\sum \limits_{\mathcal{N}=1}^{\infty}\ln
(1+e^{\frac{\mu -|E_{\mathcal{N}}|}{kT}})+ 2\ln(1+e^{\frac{\mu}{kT}})\right],
\label{SLG_OMEGA_B}\\
\Omega_{h}=\Omega_{e}(-\mu, T).
\nonumber
\end{eqnarray}
Here, we use the zero-width four-fold degenerate LL density of states as $D(E)=4\Gamma \delta(E-E_{\mathcal{N}})$, where $\Gamma=(2\pi \lambda_{B}^{2})^{-1}$ is the density of state of the single LL. Then, we assume four-fold degeneracy( spin+valley ) of  the LLs. Let us introduce the dimensionless electron energy spectrum $\varepsilon_{\mathcal{N}}=|E_{\mathcal{N}}|/kT =\sqrt{4\mathcal{N}}\vartheta$, where $\vartheta=\frac{\hbar \upsilon}{\sqrt{2}kT\lambda_{B}}$ is the reduced cyclotron energy.

Using Eqs.(\ref{NP_DEFINITION}),(\ref{SLG_OMEGA_B}) the total charge density $Q=e(N-P)$ and, finally, the quantum capacitance yields
\begin{equation}
C_{q}=\frac{e^{2}\Gamma}{kT} \left[ 4\sum \limits_{\mathcal{N}=1}^{\infty} \frac{\cosh(\varepsilon_{\mathcal{N}})\cosh \xi +1}{(\cosh(\varepsilon_{\mathcal{N}})+\cosh \xi)^{2}} + \frac{1}{\cosh^{2}(\frac{\xi}{2})} \right],
\label{CAPACITANCE_B} \\
\end{equation}
The summation term in square brackets accounts for high-index Landau levels while the second term is related to zeroth LL. The pre-factor contains the contribution to the quantum capacitance caused by single LL, $\frac{1}{4}\frac{e^{2}\Gamma}{kT}$, multiplied by the 4-fold degeneracy factor. It is to be noted also that the quantum capacitance is related to Hall conductivity as $C_{q}=\frac{eB}{c}\frac{d\sigma_{yx}}{d \mu}$.

\begin{figure}
\begin{center}
\includegraphics[scale=0.75]{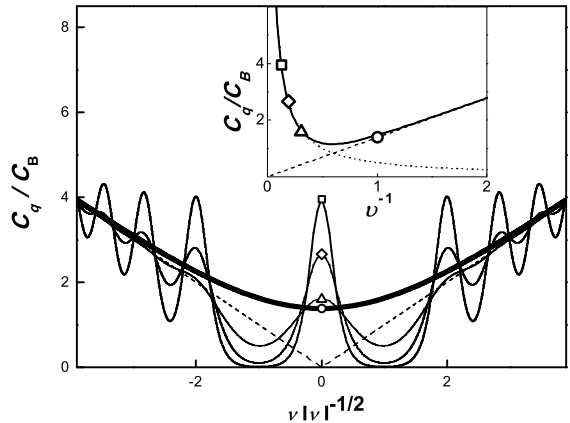} \caption[]{\label{Fig4} Dimensionless quantum capacitance $C_{q}/C_{B}$ vs square root of the filling factor at fixed B=16� and T=100,150,250K($\vartheta=7.7,5.3,3.2$). The bold line corresponds to ultrahigh-T case T=800K ($\vartheta=1$). The dashed line depicts the high-T,$\nu$ asymptote $C_{q}/C_{B}=\sqrt{|\nu|}$. Inset: temperature dependence of the quantum capacitance at DP. The low(high) temperature asymptotes are shown by the dotted(dashed) line respectively. The symbols depict DP capacitance and relate to those in the main panel.}
\end{center}
\end{figure}

At first, we intend to obtain the quantum capacitance as a function of the Fermi energy for fixed temperature and different magnitudes of the magnetic field. In order to scale the quantum capacitance in correct units, we modify the pre-factor in Eq.(\ref{CAPACITANCE_B}) as $C_{0}\vartheta^{2}/2$. The sought-for dependence $c_{q}(\xi)$  is plotted in Fig.\ref{Fig3}. In strong magnetic field the zeroth LL peak becomes more pronounced in accordance with experimental observations\cite{Ponomarenko10}. In Fig.\ref{Fig3}, inset we plot the quantum capacitance at the DP vs dimensionless magnetic field $\vartheta^{2} \sim B$. As expected, at $B=0$ the quantum capacitance is equal to zero-field value $c_{q}^{min}$ and, then grows in magnetic field. At high magnetic fields $\vartheta \gg 1$ the quantum capacitance at DP is proportional to density of states associated with zeroth LL alone. Omitting the summation, and then putting $\xi=0$ in Eq.(\ref{CAPACITANCE_B}) we obtain the high-B asymptote $c_{q}=\vartheta^{2}/2 \sim B$ shown by the dotted line in Fig.\ref{Fig3}, inset.

We now find the quantum capacitance as a function of Fermi energy for fixed magnetic field and different values of temperature. We modify the dimensional pre-factor in Eq.(\ref{CAPACITANCE_B}) as $C_{B}\vartheta/2$, where $C_{B}=\frac{2e^{2}}{\hbar \upsilon}\sqrt{\frac{\Gamma}{\pi}}$ is the T-independent capacitance unit. Then, we replace $\xi \rightarrow \nu|\nu|^{-1/2}\vartheta$ in Eq.(\ref{CAPACITANCE_B}), where $\nu=N_{0}/\Gamma$ is the conventional filling factor. The reduced quantum capacitance $C_{q}/C_{B}$ vs dimensionless variable $\nu|\nu|^{-1/2}$ for fixed magnetic field and a given temperatures in Fig.\ref{Fig4} is plotted. At low temperatures $\vartheta \gg 1$ the peaks associated with LLs are well defined. At high fillings the capacitance oscillation amplitude becomes progressively smaller compared to zeroth LL peak in accordance with experiment\cite{Ponomarenko10}. At high temperatures $\vartheta \ll 1$ the quantization of the energy spectrum becomes unimportant, hence results in $C_{q}(\nu)$-curve smoothing(see bold line in
Fig.\ref{Fig4}). At high fillings $\nu \gg \vartheta^{-2} $ the quantum capacitance follows the asymptote $C_{q}=C_{B}\sqrt{|\nu|}$ shown by dashed line in Fig.\ref{Fig4}, main panel.

Let us provide further insight into the quantum capacitance at the Dirac point. In Fig.\ref{Fig4}, inset we plot the reduced capacitance $C_{q}/C_{B}$ vs
dimensionless temperature $\vartheta^{-1} \sim T$. At low temperatures $\vartheta \gg 1$ the quantum capacitance at DP diverges as $C_{q}/C_{B}= \frac{\vartheta}{2}$ because of the trivial zero-width LL model in question. We further demonstrate that the broadening results
in finite value of the quantum capacitance at DP. In the opposite high-T case $\vartheta \ll 1$ the quantum capacitance at DP follows the trivial
zero-field asymptote $C_{q}/C_{B}=\vartheta^{-1}2\ln2$ given by Eq.(\ref{capacitance}) at $\mu=0$.

\section{ \label{sec: Influence of LL broadening on the quantum capacitance}Influence of LL broadening on the quantum capacitance}
We now improve our model regarding finite LL broadening caused by carrier scattering. For clarity, we consider the only electron part of the density of states obeyed the Gaussian form as
$D(E)=4\Gamma\sum \limits_{\mathcal{N}=0}^{\infty} \frac{1}{\sqrt{2\pi}\sigma_{0}}\exp{\frac{(E-E_{\mathcal{N}})^{2}}{2\sigma_{0}^{2}}}$,
where we assume, for a moment, the constant LLs width $\sigma_{0}$. The limit $B \rightarrow 0$ is of particular interest since the summation over the LL-index can be replaced by integration over the $\mathcal{N}$-th LL energy. After elementary transformations, the DOS yields
\begin{equation}
D(E) \vert_{B \rightarrow 0}=D_{0}(E) \frac{1}{\sqrt{2\pi}\epsilon'}\int^{\infty}_{0}t e^{-\frac{(\epsilon'-t)^{2}}{2}}dt,
\label{DOS_limitB0} \\
\end{equation}
where $\epsilon'=E/\sigma_{0}$ is the reduced energy. Note that the magnetic field is dropped out in Eq.(\ref{DOS_limitB0}). Actually, the above equation can be regarded to account the quantum states of the Dirac electrons at zero magnetic field. The question whether finite width of the states can affect DOS is fairly interesting. For high energies $\epsilon' \gg 1$ one obtains the textbook equation for zero-field DOS $D_{0}(E)$. In contrast, for low-energy case $\epsilon' \ll 1$ Eq.(\ref{DOS_limitB0}) gives a puzzling result $D(E) \vert_{B \rightarrow 0}=\frac{D_{0}(E)}{\sqrt{2\pi}\epsilon'} \sim \sigma_{0} $ implying the constant DOS. We relate the above discrepancy to our assumption of the constant broadening. Note that for $\sigma_{0}=0$ Eq.(\ref{DOS_limitB0}) yields the zero-field DOS in whole range of energies. Therefore, we conclude that $\sigma_{0} \rightarrow 0$ at $B \rightarrow 0$. This result is valid for conventional 2D electron(hole) gas as well. Furthermore, we suggest that the power function $\sigma_{0} \sim B^{k}, k \leq 1/2$ is permissible to describe LLs broadening of graphene for arbitrary magnetic field. In what follows we use the routine model for LL width as $\sigma_{0}\sim \sqrt{B}$.

\begin{figure}
\begin{center}
\includegraphics[scale=0.75]{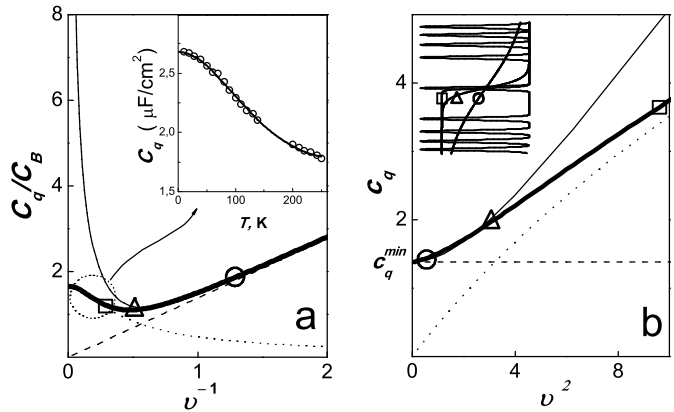} \caption[]{\label{Fig5} (a) Quantum capacitance at DP vs dimensionless temperature $\vartheta^{-1}$ at B=const. The thin (bold) line corresponds to zero-width(see Fig.\ref{Fig4}) and finite-width $\sigma_{B}=0.48$ case respectively. The low(high) temperature asymptotes
associated with finite-width capacitance are shown by the dotted(dashed) line. Inset: dimensional dependence $C_{q}(T)$ (thin line) evaluated from the
enlarged part of the bold curve in the main panel at $B$=16T and $\sigma_{0}=25$meV vs experimental data(circles) reported in Ref.\cite{Ponomarenko10};
(b) Quantum capacitance at DP vs dimensionless magnetic field $\vartheta^{2}$ at T=const. Lines notations coincide with those in panel a. Inset: the schematic view of the graphene density of states and Fermi distribution for: ($\square$)
ultra-low $\vartheta \gg 1$; ($\triangle$) intermediate $\vartheta \sim 1$; and ($\bigcirc$) high $\vartheta \ll 1$ temperature. The corresponding symbols on the
panel a,b are shown to refer these modes.}
\end{center}
\end{figure}

We now perform further insight into the quantum capacitance problem. Using the notations introduced above and the DOS in the form of Gaussian peaks the quantum capacitance yields
\begin{equation}
C_{q}=\frac{e^{2}\Gamma}{kT} \left[ \sum \limits_{\mathcal{N}=1}^{\infty} f(\mathcal{N},\xi,\sigma)+\frac{1}{2}f(0,\xi,\sigma) \right],
\label{CAPACITANCE_B+} \\
\end{equation}
where $f(\mathcal{N},\xi,\sigma)=\frac{4}{\sqrt{2\pi}\sigma}\int \limits_{-\infty}^{\infty} \frac{\cosh \varepsilon \cosh \xi +1}{(\cosh \varepsilon+\cosh \xi)^{2}} e^{\frac{(\epsilon-\epsilon_{\mathcal{N}})^{2}}{2\sigma^{2}}} d\epsilon$ is a function  accounting the LL broadening and temperature effects combined. Then, $\sigma=\sigma_{0}/kT=\sigma_{B}\vartheta$ is the dimensionless LL width, $\sigma_{B}=\frac{\sigma_{0}\lambda_{B}\sqrt{2}}{\hbar \upsilon}$ is a constant ratio  of the LL width to cyclotron energy. Our previous result of zero-width LLs remains valid at $\sigma \ll 1$ when Eqs.(\ref{CAPACITANCE_B+})(\ref{CAPACITANCE_B}) coincide. Hereafter, we will be mostly interested in the opposite wide-width LLs case $\sigma \gg 1$.

We refer first to the case of magnetic field kept constant. The inset in Fig.\ref{Fig5}b depicts the actual view of separated LLs($\sigma_{B}<1$) and the
sketch of the Fermi distribution associated with low($\vartheta \gg 1$), intermediate($\vartheta \sim 1$) and high $\vartheta \ll 1$ temperatures.
One expects that the role playing by LL broadening is somewhat similar to that caused by temperature(see Fig.\ref{Fig4}). Namely, at low enough temperatures the bigger the LL width, the smoother the capacitance oscillations. In order to illustrate the importance of LLs broadening, in Fig.\ref{Fig5},a we compare the
quantum capacitance at DP for both the zero and finite LLs width cases. At low temperatures $\vartheta \gg 1$ the quantum capacitance is described by Eq.(\ref{CAPACITANCE_B+}) with the high-index LLs neglected, i.e $C_{q}/C_{B}=\frac{\vartheta}{4}f(0,0,\sigma)$. Contrary to divergent behavior of the capacitance found earlier for zero-width case at $T \rightarrow 0$, in presence of broadening the capacitance approaches a finite value $C_{q}/C_{B}=\sqrt{\frac{2}{\pi}}\frac{1}{\sigma_{B}}$. As an example, in Fig.\ref{Fig5},a we plot the DP capacitance at $\sigma_{B}=0.48$. For intermediate temperatures $\vartheta \leq 1$ the influence of LLs broadening on the quantum capacitance becomes less important compared to role playing by temperature. For ultra-high temperatures $\vartheta \ll 1$ the LL quantization can be disregarded completely, therefore the quantum capacitance follows familiar zero-field asymptote $C_{q}/C_{B}=\vartheta^{-1}2\ln2$.

In order to check the relevance of our approach, in Fig.\ref{Fig5}a, inset we fit the experimental data\cite{Ponomarenko10} using dimensional form of the dependence shown by the bold line in Fig.\ref{Fig5}a. The extracted value of the LL width $\sigma_{0}=25$meV at $B=16$T is close to that $\sim 30$meV reported in Ref.\cite{Ponomarenko10}.

Let us now examine the quantum capacitance at DP for fixed temperature and varied magnetic field. One can easily re-plot the curves shown in Fig.\ref{Fig5},a
in terms of dimensionless capacitance $c_{q}$ and magnetic field $\vartheta^{2}$. The result is presented in Fig.\ref{Fig5},b. The zero and finite width cases still differ one from another at low-temperatures and(or) high-magnetic fields $\vartheta \gg 1$. Our model implying LLs width
vanishing at $B \rightarrow 0$ gives the expected value of the zero-field quantum capacitance $c_{q}^{min}$. Experimentally, the latter remains, however, unchanged\cite{Ponomarenko10} within wide range of temperatures because of the electron-hole plasma puddles.

It is instructive to discuss the constant width model made use in Ref.\cite{Ponomarenko10}. Surprisingly, the zero-field quantum capacitance at DP was claimed to depend on LL width in absence of the magnetic field itself. We attribute this puzzling result to unavoidable overlapping of the LLs at $B \rightarrow 0$ under the condition of the constant broadening. Thus, the applicability of the constant broadening model is doubtfully.

\section{ \label{sec: Quantum capacitance in ultra quantum limit}Quantum capacitance in ultra quantum limit}
\begin{figure}
\begin{center}
\includegraphics[scale=0.75]{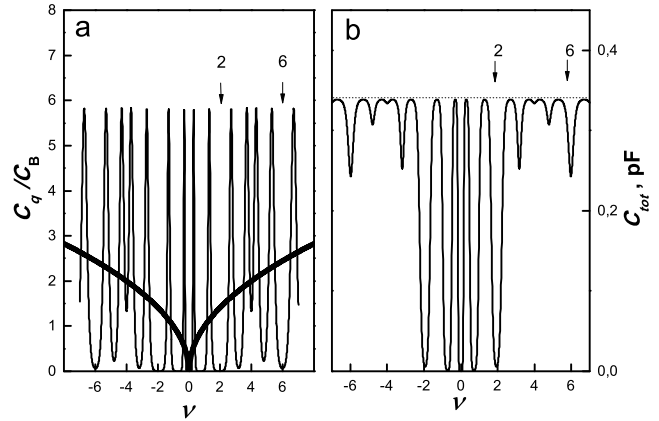} \caption[]{\label{Fig6} (a) Dimensionless quantum capacitance $C_{q}/C_{B}$ vs filling factor for splitted
LLs(non-broadened) at $B$=15� and T=20K. The bold line corresponds to degenerated carriers at $T,B$=0. (b) The total capacitance of the graphene
sample taking into account the geometric capacitance $C$=0.34pF\cite{Yu13}. }
\end{center}
\end{figure}

Recently, the measurements of the graphene quantum capacitance became available\cite{Yu13} for large-area high-quality samples placed in quantizing magnetic fields. In contrast to previous data\cite{Ponomarenko10}, at ultra-low temperatures the splitting of four-fold degenerate LLs becomes clearly visible. The most pronounced splitting was observed for zeroth LL. We verify that the splitting magnitude of zeroth LL correlates with that extracted\cite{Cheremisin11} from transport measurements data. Moreover, the authors of Ref.\cite{Yu13} suggest that the high-order LLs splitting strengths are comparable with those for zeroth LL. Let us compare the experimental results\cite{Yu13} with those predicted by theory. Using the experimental data for $B$=15T we are able to deduce the fillings of the zeroth LL sub-levels as $\pm 0.31; \pm 1.28$. Following the arguments put forward in Ref.\cite{Yu13} we further assume the same splitting values for higher LLs. For simplicity, we neglect the LLs broadening, thus use Eq.(\ref{CAPACITANCE_B}) modified with respect to actual splitting of LLs spectrum. In Fig.\ref{Fig6},a we plot the dimensionless quantum capacitance as a function of the filling factor. In contrast to experiment we use somewhat higher temperature $T$=20K due to restrictions of our numerical method. The overall behavior of the quantum capacitance in Fig.\ref{Fig6} resembles that represented in Fig.\ref{Fig6} for unperturbed energy spectrum. One can easily find the total capacitance of the graphene device taking into account the geometric capacitance contribution. For $B$=15T, velocity $\upsilon=1.15 \times 10^{8}$ cm/s\cite{Yu13} we estimate $C_{B}=1.43\mu$F/cm$^{2}$. Therefore, for actual sample area $S\sim 10^{3}$ $\mu$m$^{2}$ we obtain the device quantum capacitance as $C_{B}S=14.3$pF. The latter is 40 times greater than the reported geometric capacitance $C=0.34$pF. Finally, in Fig.\ref{Fig6},b we plot the total capacitance $C_{tot}$ of the graphene sample. The main difference with respect to experimental data\cite{Yu13} concerns the ratio between the dip strength at $\nu=2$ and those related to gaps in zeroth LL split spectrum.

In conclusion, we calculated the quantum capacitance of the monolayer graphene for arbitrary magnetic field, temperature, LL broadening and
realistic splitting of LL spectrum. Magnetic field measurements of the capacitance\cite{Ponomarenko10}, \cite{Yu13} carried out at
high- and low- temperature modes are consistent with our approach.

The author wish to thank Dr. A.P.Dmitriev for helpful comments.

\end{document}